\documentclass[
    amsmath,
    amssymb,
    reprint,
    aps,
    prx,
    superscriptaddress,
    longbibliography
]{revtex4-1}

\usepackage[utf8]{inputenc}
\usepackage[T1]{fontenc}
\usepackage{physics}
\usepackage{nicefrac}
\usepackage{bbm, bm}
\usepackage{makecell}
\usepackage{multirow}
\usepackage{booktabs}
\usepackage{overpic}
\usepackage{tikz}
\usepackage{pgfplots}
\pgfplotsset{compat=1.18}
\usepackage[dvipsnames]{xcolor}
\usepackage{ragged2e}

\usepackage[breaklinks=true, colorlinks]{hyperref}
\hypersetup{linkcolor=blue, citecolor=blue, urlcolor=MidnightBlue}
\usepackage{cleveref}

\begin{document}

\title{A minimal mechanism to generate long timescales 
without fine tuning}

\author{Kathryn McClain*}
\affiliation{Center for Neural Science, NYU} 
\affiliation{Center for Soft Matter Research, Department of Physics, NYU} 

\author{Shivang Rawat*}
\affiliation{Center for Soft Matter Research, Department of Physics, NYU}
\affiliation{Courant Institute of Mathematical Sciences, NYU}

\author{Mia Morrell}
\affiliation{Center for Neural Science, NYU} 
\affiliation{Center for Soft Matter Research, Department of Physics, NYU} 

\author{Stefano Martiniani}
\affiliation{Center for Neural Science, NYU} 
\affiliation{Center for Soft Matter Research, Department of Physics, NYU} 
\affiliation{Courant Institute of Mathematical Sciences, NYU} 
\affiliation{Simons Center for Computational Physical Chemistry, Department of Chemistry, NYU} 

\author{David J. Heeger}
\affiliation{Department of Psychology and Center for Neural Science, NYU}

\author{Flaviano Morone}
\affiliation{Center for Neural Science, NYU} 
\affiliation{Center for Soft Matter Research, Department of Physics, NYU}

\begin{abstract}
Long timescales in brain dynamics give 
rise to power law correlations measured 
in experiments. A simple linear recurrent 
neural network model can reproduce this 
power law behavior, but the recurrent 
interaction strength needs to be fine 
tuned in order to sit at the edge of 
stability. We show that by adding a 
dynamical recurrent feedback gain to the 
simple linear model removes the need for 
fine tuning, because the gain self-organizes 
so that the spectrum of the effective 
recurrent matrix is always gapless with 
the top eigenvalue landing on the stability 
edge for any recurrent strength. As a 
consequence, long timescales and scale free 
correlations arise generically, as we 
demonstrate both analytically and numerically. 
\end{abstract}

\maketitle

%\onecolumngrid
\emph{Introduction}
Neurons operate on timescales of milliseconds, 
yet the brain processes information over 
timescales many orders of magnitude 
longer~\cite{bernacchia2011, zeisler2025, runyan2017, zeraati2026, huang2017}, supporting behaviors 
that unfold over seconds, hours, and years, 
from motor learning to working memory. 

A minimal model for this separation of timescales 
is a linear stochastic network of $N$ recurrently 
interacting neurons
\begin{equation}
\tau_x\dot{x}_i = -x_i + \sum_{j=1}^NM_{ij}x_j + \xi_i\ ,
\label{eq:linearmodel}
\end{equation}
whose modes relax with effective timescales 
$\tau_\lambda = \tau_x/(1-\lambda)$, one for 
each  eigenvalue $\lambda$ of the recurrent 
matrix $M$~\cite{chen2024}. 
At first sight this is all we need, since a 
large network naturally displays a whole 
spectrum of timescales. This spectrum, however, 
is generically a spectrum of short timescales. 
Any mode with $\lambda$ well below 1 relaxes 
within a few intrinsic time constants $\tau_x$, 
and longer timescales appear only for eigenvalues approaching the stability edge $\lambda\to1^-$, 
where $\tau_\lambda$ diverges. 
Everything therefore hinges on the top of 
the spectrum. If the eigenvalue density 
$\rho(\lambda)$ has a gap below 1, all 
correlations induced by the white noise 
$\xi(t)$ decay exponentially on the timescale 
of the slowest mode. If instead the spectrum 
is gapless, with 
$\rho(\lambda)\propto(1-\lambda)^\alpha$ near 
the edge, the correlations decay algebraically 
as $C(t)\sim t^{-\alpha}$, with no characteristic scale~\cite{chen2024}. 
Reaching the edge, however, requires fine tuning. 
Nothing in a generic network anchors the largest 
eigenvalue at the stability edge, and the two 
coincide only for special values of the synaptic couplings~\cite{chen2024}. 
Power law correlations of neural activity 
are measured consistently across individuals 
and species~\cite{meshulam2019, he2014}, 
but development, plasticity, and evolution 
guarantee that no two brains share the same 
connectivity. 
Within the linear model~\eqref{eq:linearmodel} 
this is a conundrum: 
scale-free correlations exist only for a 
single special set of synaptic couplings, 
whereas experiments find them in every brain, 
each wired differently. A version of this problem 
was recognized already thirty years ago in 
the context of neural integrators~\cite{seung1996, major2004}. 
The recurrent interactions alone cannot be the 
solution. Some mechanism beyond the couplings 
themselves must place the network at the stability 
edge and hold it there.

A network at this edge, with diverging timescales 
and scale-free correlations, is a system at 
criticality. In the brain, criticality has been 
associated with optimal information
transmission~\cite{beggs2003, plenz2021, ShewPlenz2013}, 
maximal dynamic range in response to
stimuli~\cite{gautam2015,kinouchi2006}, and slow 
collective modes supporting working 
memory~\cite{bernacchia2011, major2004,ganguli2008}, 
while its loss has been connected to pathologies 
such as epilepsy~\cite{zimmern2020, Meisel2012}.

Several mechanisms have been proposed to reach 
and hold the edge, differing in which ingredient 
of the network they modify. A first class shapes 
the statistics of the connectivity so that the 
spectrum is gapless by construction, through 
special ensembles~\cite{chen2024}, network
topology~\cite{zenari2026}, or the structure of 
excitation and inhibition~\cite{rajan2006}. 
However, this displaces the problem rather than 
solving it, in that the fine-tuning of the coupling 
is traded for a special choice of the ensemble. 
A second class lets the weights themselves evolve: 
criticality is viewed as a target state of the 
network, with synaptic plasticity slowly correcting 
deviations from it~\cite{bienenstock1982, hengen2025,yellin2025,meisel2017}. 
Notably, anti-Hebbian rules can self-tune a network 
to a critical state~\cite{Magnasco2009}. Plasticity, 
however, acts on the $N^2$ synaptic weights and on 
timescales of hours to days, leaving open how 
scale-free correlations persist during ongoing 
activity. Closer to our work, multiplicative 
gating achieves marginal stability without 
fine-tuned weights~\cite{krishnamurthy2022, can2025}, 
but at the price of a high-dimensional 
architecture with $O(N)$ gating variables, 
one per neuron. 

Here we show that a single additional 
unit suffices, one that dynamically pools the total 
activity of the network and suppresses the recurrent 
gain in proportion. This is not an exotic circuit. 
It is precisely the pooled feedback known as 
normalization, a canonical computation observed 
throughout the cortex and in other brain areas~\cite{Heeger1992, CarandiniHeeger2012,heeger2019oscillatory}. 
Mechanistically, interneurons are the key 
mediators of normalization~\cite{atallah2012}, 
and are implicated in maintaining criticality in 
cortical networks~\cite{ma2019}. Functionally, 
normalization maximizes sensitivity by aligning 
the range of neural responses to the range of 
sensory inputs~\cite{CarandiniHeeger2012}, 
and this same dynamic range optimization is 
observed near criticality in the sensory 
cortex~\cite{gautam2015}. 

Recently, normalization has been shown to 
stabilize recurrent networks well beyond the 
stability limit of linear models, with slow 
dynamics emerging over a broad range of coupling 
strengths~\cite{morone2026stabilization}. Here 
we show that this mechanism does more than stabilize: 
it pins the largest eigenvalue at the stability 
edge, for any interaction strength and with no 
adjustment of the synaptic weights, converting 
the critical point of the linear model into a 
critical phase, with scale-free correlations 
throughout. 

\medskip

\emph{Setup --}~Our model consists of $N$ 
recurrently connected neurons $x_i$ coupled 
to a single additional unit $a$, which pools 
the total activity of the network, 
$||x||^2=\sum_ix_i^2$, and feeds back by scaling 
every recurrent synapse by the common factor 
$(1-a)$. The dynamics is 
\begin{equation}
\begin{aligned}
\tau_x\dot{x}_i &= -x_i + (1-a)\sum_jM_{ij}x_j + \xi_i\ ,\\
\tau_a \dot{a}  &= -a + \sigma^2 + a\sum_i x_i^2\ .
\end{aligned}
\label{eq:model}
\end{equation}
Note that the gain unit is blind to the 
individual weights $M_{ij}$. It senses only the overall level of 
activity and responds by uniformly suppressing 
the recurrent couplings, akin to the pooled 
divisive feedback of normalization~\cite{Heeger1992, CarandiniHeeger2012, heeger2019oscillatory}. 
The parameter $0<\sigma^2<1$ is a small floor 
for the gain. When $a$ approaches zero, the 
second equation reduces to $\tau_a\dot{a}\approx\sigma^2$, 
so a positive initial gain remains positive 
at all times. 
\begin{figure}[h!]
	\centering
	\includegraphics[width=0.48\textwidth]{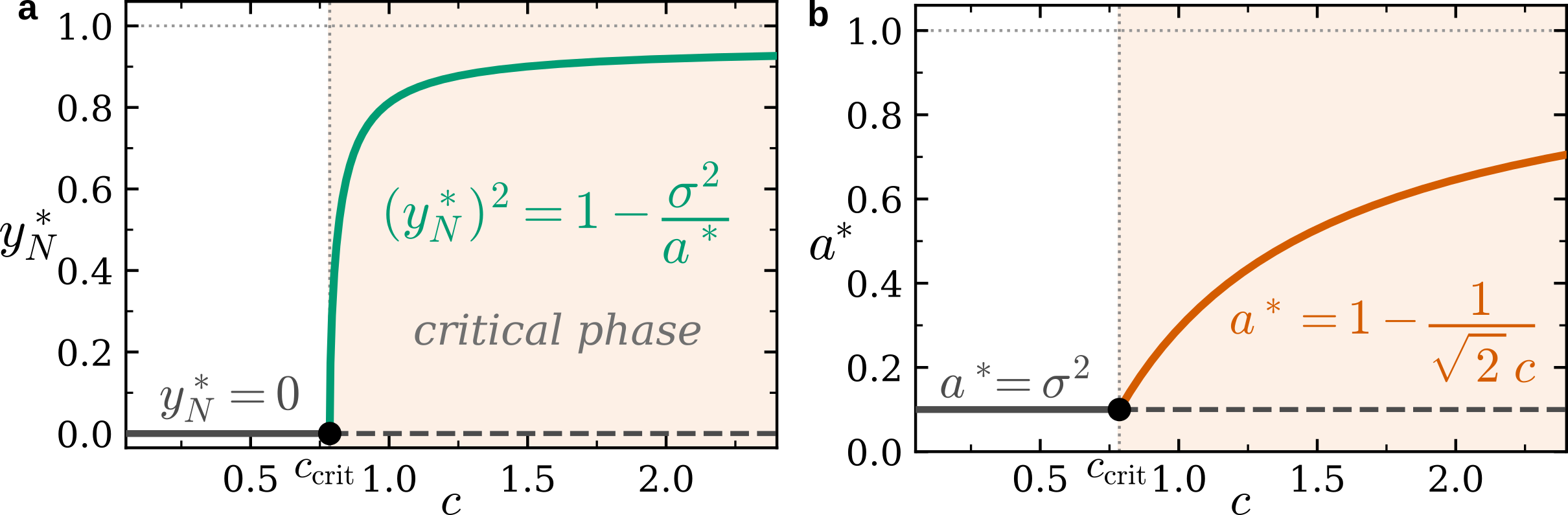}
	\caption{
    %\raggedright
    \justifying
    \textbf{Fixed points of the dynamics~\eqref{eq:model_diagonal} 
    versus interaction strength $c$.}
    {\bf a}, Activity $y_N^*$ of the top mode and {\bf b}, 
    gain $a^*$. Solid lines denote the stable fixed points 
    and the dashed lines the unstable ones. 
    At $c_{crit}=1/[\sqrt{2}(1-\sigma^2)]$ (black dot) 
    the trivial fixed point loses stability and the 
    active fixed point Eq.~\eqref{eq:activeFP} emerges 
    continuously. In the whole shaded region the gain 
    settles at the value $a^*=1-1/(\sqrt{2} c)$ that 
    places the top eigenvalue of $(1-a^*)M$ exactly at 
    the stability edge, so that $c> c_{crit}$ is a critical 
    phase with no tuning of $c$ required. 
    }
    \label{fig:fig1}
\end{figure}

The recurrent matrix $M$ may be drawn, for example, 
from the Gaussian Orthogonal Ensemble (GOE): the 
entries are independent Gaussian variables, symmetric 
$M_{ij}=M_{ji}$, with zero mean and variance 
$c^2/N$ on the diagonal and $c^2/2N$ off the 
diagonal, following the notation of Ref.~\cite{chen2024}. 
The single parameter $c$ sets the strength 
of the recurrent interactions, and for large 
$N$ the eigenvalues of $M$ obey the Wigner 
semicircle law on the interval 
$[-\sqrt 2 c, \sqrt 2 c]$. 
The noise $\xi_i$ is weak and white with zero 
mean and $\langle \xi_i(t) \xi_j(t')\rangle = 2D\delta_{ij}\delta(t-t')$. 
For $a=0$ the model reduces to the linear network  
of Chen and Bialek~\cite{chen2024}. 
The gain dynamics is the crucial new ingredient. 
Because $M$ is symmetric, it is diagonalized by 
an orthogonal transformation, $M=O\Lambda O^T$, 
and setting $y=O^Tx$ gives the equivalent system 
\begin{equation}
\begin{aligned}
\tau_x\dot{y}_i &= -\gamma_i(a)y_i + \tilde{\xi}_i\ ,\\
\tau_a \dot{a}  &= -a + \sigma^2 + a||y||^2\ ,
\end{aligned}
\label{eq:model_diagonal}
\end{equation}
where $\tilde{\xi}=O^T\xi$ has the same 
statistics as $\xi$, and each mode relaxes 
with the bare decay rate 
\begin{equation}
\gamma_i(a)\equiv 1-(1-a)\lambda_i\ . 
\end{equation}
Hereafter we set $\tau_x=\tau_a=1$, since 
none of the results below depend on this choice. 
At fixed $a$, the modes behave exactly as in 
a linear model with the effective recurrent matrix 
$(1-a)M$. The gain, however, is not fixed, 
but driven by the total activity and any 
change of the gain moves all decay rates 
simultaneously. This closed loop between 
activity and gain distinguishes our system in Eq.~\eqref{eq:model_diagonal} from any linear 
model with a static gain. 

%\medskip

\emph{Results --}~At zero noise, the 
system~\eqref{eq:model_diagonal} admits 
two types of fixed points. The trivial 
fixed point has no activity, $y_i^*=0$ 
for all $i$, and $a^*=\sigma^2$. Linearizing 
around it, each mode relaxes at rate 
$\gamma_i(\sigma^2)=1-(1-\sigma^2)\lambda_i$, 
which is positive for all modes provided 
$(1-\sigma^2)\lambda_N<1$. For the GOE, 
where $\lambda_N\to\sqrt2c$, this gives
\begin{equation}
 \textrm{trivial FP stable if}\ \ \ 
c < c_{crit}=\frac{1}{\sqrt{2}(1-\sigma^2)}\ .
\label{eq_critical_connectivity}
\end{equation}
We use $\sigma^2=0.1$ throughout, so 
$c_{crit}\approx1/\sqrt{2}$. In this regime the 
model behaves like the linear model~\cite{chen2024}: 
correlations decay exponentially for 
$c < c_{crit}$, and algebraically, as $t^{-1/2}$ 
with a cutoff $t_{max}\propto N^{2/3}$, exactly 
at $c_{crit}$. We have gained nothing yet, 
since criticality still requires tuning $c$. 

For $c > c_{crit}$ the picture changes completely. 
The trivial fixed point is unstable, since the 
top mode grows, but the growing activity drives 
the gain up, which reduces the effective coupling 
$(1-a)M$ until growth stops. The result is an 
active fixed point, which we now construct. Any 
fixed point with nonzero activity must satisfy 
$\gamma_i(a^*)y_i=0$ for every mode. Since the 
rates $\gamma_i(a^*)$ are all distinct for a 
non-degenerate spectrum, at most one mode can be 
active. Which mode is selected? For any eigenmode 
$k$ one can set $y_i^*=0$ for $i\neq k$ and choose 
the gain to make its rate vanish as $a^*=1-1/\lambda_k$. 
At such a candidate fixed point, however, the 
remaining modes relax at rates 
$\gamma_i(a^*) = 1 - \lambda_i/\lambda_k$, which 
are negative for every $\lambda_i>\lambda_k$. 
Therefore, each eigenvalue above $\lambda_k$ is 
an unstable direction. Only the choice $k=N$ leaves 
no eigenvalue above, hence no unstable direction. 
Stability thus selects the top mode automatically. 
The active fixed point is therefore unique (up to 
the sign of $y_N^*$) and reads
\begin{equation}
\begin{aligned}
(y_N^*)^2 &= 1 - \frac{\sigma^2}{a^*}\ ,
\qquad y_i^* = 0 \quad (i < N)\ ,\\
a^* &= 1 - \frac{1}{\lambda_N}\ .
\end{aligned}
\label{eq:activeFP}
\end{equation}
The fixed point exists when $(y_N^*)^2>0$, 
i.e. $a^*>\sigma^2$, or equivalently 
$c>c_{crit}$, so the active state is born 
exactly where the trivial state loses 
stability, as seen in Fig.~\ref{fig:fig1}. 
Note also that $a^*$ lies strictly between 
$\sigma^2$ and $1$ for any $c>c_{crit}$, 
approaching $1$ only as $c\to\infty$ 
(see Fig.~\ref{fig:fig1}b). The gain thus 
suppresses the recurrent drive without 
ever reversing its sign, and $(1-a)$ acts 
as a stabilizing negative feedback at 
arbitrarily strong coupling. 
\begin{figure}[h!]
	\centering
        \vspace{-2mm}
	\includegraphics[width=0.36\textwidth]{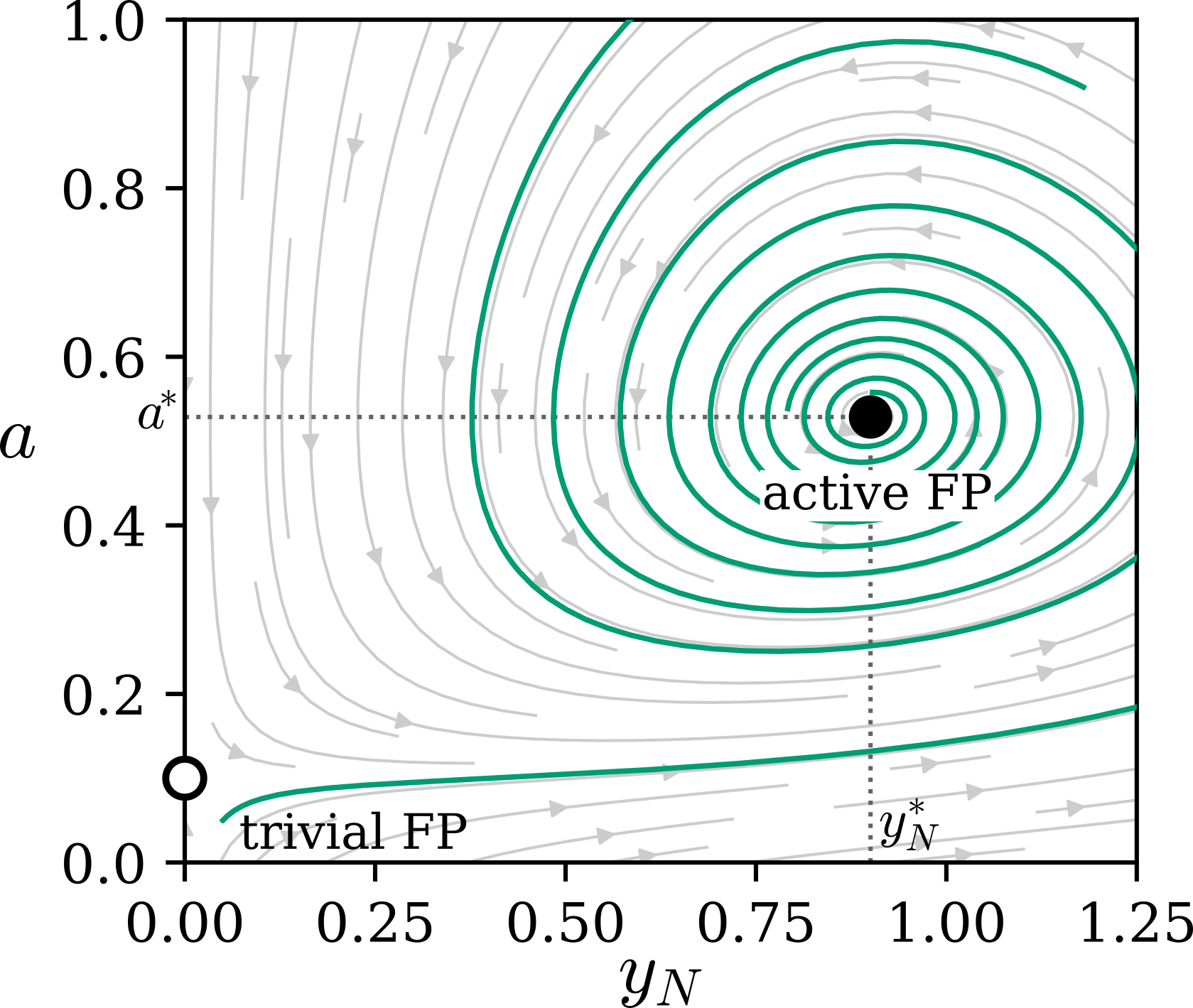}
	\caption{
    %\raggedright
    \justifying
    \textbf{Relaxation to the active fixed point}. 
    Deterministic flow in the $(y_N,a)$ plane for 
    $c=1.5$ and $\sigma^2=0.1$ (only $y_N>0$ shown). 
    The trivial fixed point (open circle) is unstable, 
    and trajectories spiral into the active fixed 
    point (filled circle) at $(y_N^*,a^*)$, with 
    frequency $\omega\approx\sqrt{\det J_2}$, 
    due to the complex eigenvalues 
    $\mu_\pm=\beta\pm i\omega$ of the block 
    $J_2$ in Eq.~\eqref{eq:J2block}. 
    }
    \label{fig:fig2}
    \vspace{-3mm}
\end{figure}

Equation~\eqref{eq:activeFP} is the core of 
the mechanism. The condition $\gamma_N(a^*)=0$ 
means $(1-a^*)\lambda_N=1$, which shows how 
the gain settles at precisely the value that 
places the top eigenvalue of the effective 
recurrent matrix on the stability edge, for 
any $c > c_{crit}$. Moreover, the gain locks 
onto the realized $\lambda_N$ of the particular 
draw of $M$, not its ensemble average, meaning 
that the self-organization is robust not only 
to the interaction strength but also to 
sample-to-sample fluctuations of the connectivity. 

We now linearize Eq.~\eqref{eq:model_diagonal} 
around the active fixed point. The Jacobian 
inherits a simple structure from the fixed 
point itself. The gain couples to the modes 
only through $||y||^2$, so its linearization 
involves $\partial\dot{a}/\partial y_i = 2a^*y_i^*$, 
which vanishes for every transverse 
mode because $y_i^*=0$ for $i<N$. 
At linear order the gain talks only to the 
active mode, so the Jacobian splits exactly 
into a diagonal block of $N-1$ transverse 
modes and a $2\times2$ block coupling the 
pair $(y_N,a)$. The transverse eigenvalues 
are the rates already obtained in the argument 
for the mode selection, evaluated at $k=N$, 
\begin{equation}
\mu_i = -\gamma_i(a^*) = -\Big(1-\frac{\lambda_i}{\lambda_N}\Big) < 0\ , \ \ i=1,...,N-1\ .
\label{eq:transverserates}
\end{equation}
Since the interaction strength $c$ sets only 
the overall scale of the spectrum, it cancels 
from the ratios $\lambda_i/\lambda_N$ and hence 
from the relaxation rates $\mu_i$. The gain has erased 
precisely the one parameter that required tuning. 
All transverse rates are strictly negative 
at finite $N$, so the fixed point is stable 
in every transverse direction. The $2\times2$ 
block reads
\begin{equation}
J_2 = \begin{pmatrix}
    0 & -y_N^*\lambda_{N}\\
    2a^*y_N^* & -\sigma^2/a^*
\end{pmatrix}\ .
\label{eq:J2block}
\end{equation}
Both eigenvalues of $J_2$ have negative real 
part for any $c>c_{crit}$, meaning that the 
active fixed point, whenever it exists, is 
linearly stable. 
The structure of this block deserves emphasis. 
The bare decay rate of the active mode is marginal 
by construction, $\gamma_N(a^*)=0$, which is 
the zero in the upper-left corner of $J_2$, yet 
the mode is not marginal since its coupling 
to the gain renormalizes the decay rate to a strictly 
negative value. The feedback loop that pins 
the recurrent spectrum to the edge thus also 
sequesters the marginal mode, pairing it with 
the gain into a strictly stable subsystem, 
whose damped relaxation toward the fixed point 
is shown in Fig.~\ref{fig:fig2}.

Two spectra should be distinguished here. The 
first is that of the effective recurrent matrix 
$(1-a^*)M$, whose eigenvalues are $(1-a^*)\lambda_i$. 
Its top eigenvalue equals $1$ exactly, so the 
effective matrix sits precisely at the stability 
edge. The second is that of the Jacobian, which 
governs the relaxation around the fixed point 
and consists of the $N-1$ transverse rates $\mu_i$ 
together with the pair $\mu_{\pm}$ from $J_2$. 
This spectrum contains no marginal eigenvalue, 
since at any finite $N$ all its eigenvalues have 
strictly negative real part. What survives of the 
marginality is the accumulation of the transverse 
rates at zero, with a spectral gap that closes as 
$N\to\infty$. Criticality thus originates from the 
vanishing of the spectral gap of the Jacobian. 

\begin{figure}[ht!]
	\centering
    \vspace{3mm}
	\includegraphics[width=0.36\textwidth]{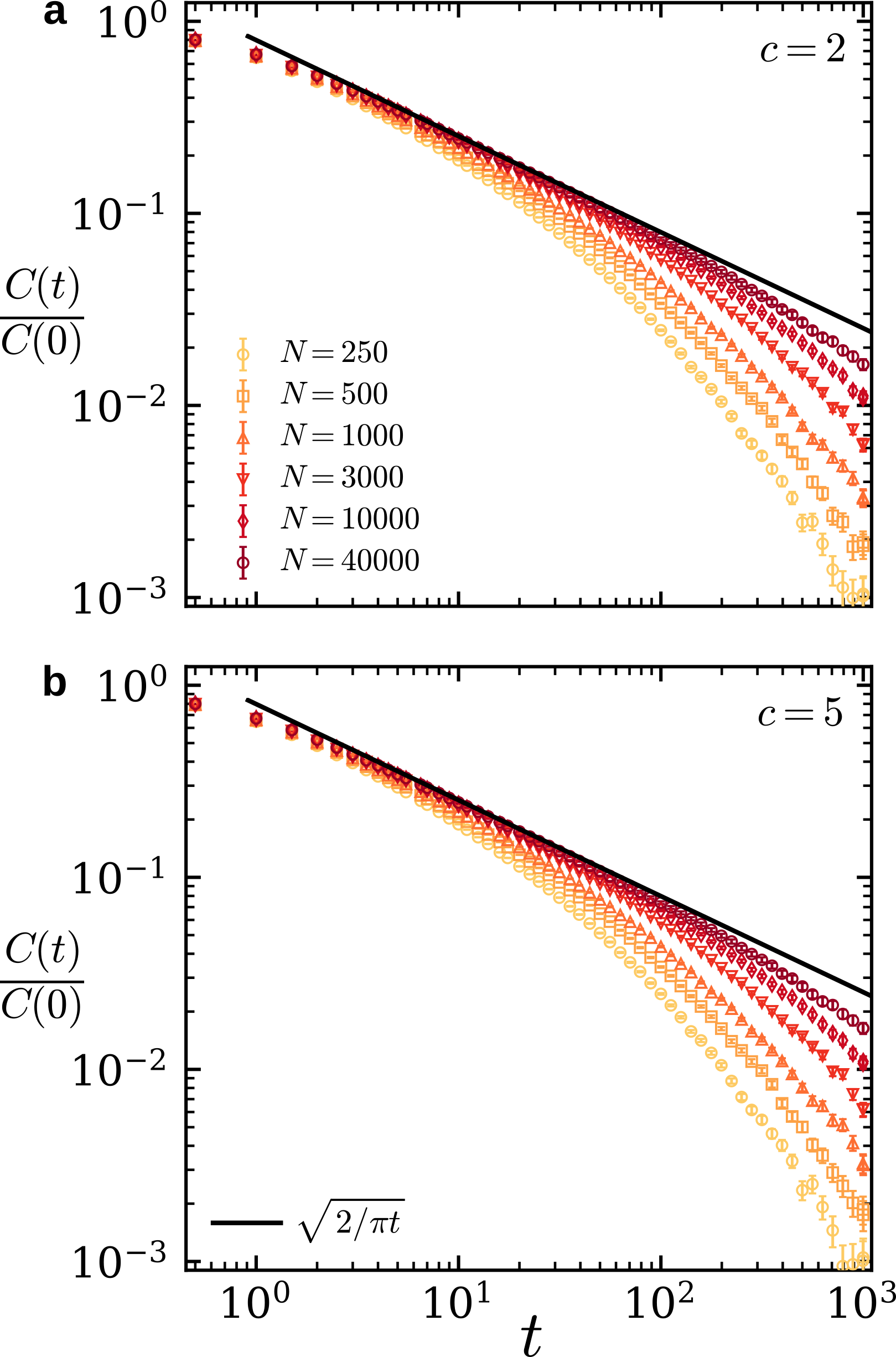}
	\caption{
    %\raggedright
    \justifying
    \textbf{Scale-free correlations throughout the 
    critical phase.} Normalized correlation function 
    of the transverse modes for {\bf a}, $c=2$ and 
    {\bf b}, $c=5$, at system sizes from $N=250$ to 
    $N=40000$, obtained by numerical integration of 
    Eq.~\eqref{eq:model_diagonal} with $\sigma^2=0.1$ 
    and noise amplitude $D=0.01/N$, averaged over many 
    realizations of $M$ (error bars are s.e.m.). The 
    black line is the prediction $\sqrt{2/\pi t}$ of 
    Eq.~\eqref{eq:powerlaw}, not a fit. The data follow 
    the power law up to the cutoff $t_{max}\propto N^{2/3}$, 
    after which the decay is exponential. Note that the 
    two panels are indistinguishable, with no dependence  
    on $c$ in the scale-free regime.
    }
    \label{fig:fig3}
    \vspace{-4mm}
\end{figure}

Adding noise unveils our central result. Around 
the active fixed point, the linearized stochastic 
dynamics is a set of Ornstein-Uhlenbeck 
(OU) processes where the $N-1$ transverse 
fluctuations $\delta y_i$ relax independently at 
rates $|\mu_i|=1-\lambda_i/\lambda_N$, while the 
pair $(\delta y_N, \delta a)$ is a two-dimensional 
OU process governed by $J_2$. We consider the 
average autocorrelation of the neural activity
\begin{equation}
\begin{aligned}
C(t) &= \frac{1}{N}\sum_{i=1}^N \langle \delta x_i(0)\, \delta x_i(t) \rangle = 
\frac{1}{N}\sum_{i=1}^N \langle \delta y_i(0)\, \delta y_i(t) \rangle\ ,
%\frac{1}{N}\sum_{i=1}^N \frac{1}{|\mu_i|}e^{-|\mu_i||t|}\ .
\end{aligned}
\end{equation}
where the second equality follows from the 
orthogonality of $O$. The sequestered pair 
contributes one term out of $N$, decaying 
at a finite rate, so it is negligible 
at long times. The long-time behavior comes 
entirely from the transverse modes, each 
contributing a standard OU correlation, 
thus giving 
\begin{equation}
C(t) = \frac{1}{N}\sum_{i=1}^{N-1} 
\frac{D}{|\mu_i|} e^{-|\mu_i||t|}\ . 
\end{equation}
For large $N$ the sum becomes an integral over 
the density of relaxation rates $\rho(\mu)$, 
obtained from the semicircle law by the change 
of variables $\mu=1-\lambda/\lambda_N$ (removing 
the single eigenvalue $\lambda_N$ does not alter 
the limiting density), which gives 
$\rho(\mu)=\frac{2}{\pi}\sqrt{\mu(2-\mu)}$,  
with $\mu\in[0,2]$. Every trace of $c$ has 
canceled, and $\rho(\mu)$ is the same for 
all $c>c_{crit}$. The correlation function, 
normalized by its equal-time value $C(0)=2D$, 
is then 
\begin{equation}
\frac{C(t)}{C(0)} = \frac{1}{2}\int_0^2 d\mu\, 
\frac{\rho(\mu)}{\mu}\, e^{-\mu |t|} \approx 
\sqrt{\frac{2}{\pi|t|}}\ ,\qquad |t|\gg1\ ,
\label{eq:powerlaw}
\end{equation}
where the long-time behavior is governed by the 
$\sqrt{\mu}$ edge of $\rho(\mu)$. Both the coupling 
$c$ and the noise amplitude $D$ have dropped out of 
Eq.~\eqref{eq:powerlaw}, which is therefore a 
parameter-free prediction confirmed by the simulations 
for $c=2$ and $c=5$ in Fig.~\ref{fig:fig3}. 
At finite $N$ the power law is cut off. Indeed, 
since the eigenvalue spacing at the GOE edge scales 
as $\lambda_N-\lambda_{N-1}\propto N^{-2/3}$, the 
scale-free decay extends up to $t_{max}\propto N^{2/3}$, 
beyond which $C(t)$ decays exponentially, as 
visible in Fig.~\ref{fig:fig3} for the smaller 
sizes.

%\medskip

\emph{Discussion --}~In the linear model~\cite{chen2024}, 
the $t^{-1/2}$ decay with $N^{2/3}$ cutoff exists 
only at the isolated point $c=c_{crit}$. The gain 
converts this critical point into a critical phase, 
with the same exponent and cutoff scaling for every 
$c$ above the threshold, resolving the fine-tuning 
conundrum.

In the critical phase, the active fixed point lies 
along a single eigenmode of $M$, a pattern broadly 
distributed across the neurons, comparable to the 
functional attractors observed in the brain~\cite{seung1996}. 
Around it, the continuum of slow modes produces 
high-dimensional fluctuations for any $c>c_{crit}$, 
while the fixed point remains stable, because 
$\lambda_N$ stays pinned by the gain however 
large $c$ grows. A zero-dimensional attractor thus 
coexists with high-dimensional critical dynamics, 
which may reconcile the low-dimensional attractors 
found throughout the brain~\cite{perich2025} with 
the complex dynamics observed at criticality~\cite{fosque2025}. 
In our model, stability and long timescales coexist.

Our mechanism is not tied to the GOE. In fact, for 
any symmetric ensemble whose spectral density vanishes 
at the edge as $(\lambda_{edge}-\lambda)^\alpha$, 
the same construction yields a critical phase with 
$C(t)\sim t^{-\alpha}$ and cutoff 
$t_{max}\sim N^{1/(1+\alpha)}$. Normalization thus 
self-organizes the network to the edge, while the 
ensemble sets the exponent. Since $\alpha=1/2$ is 
the generic value for random matrices, any departure 
from $t^{-1/2}$ would require connectivity with 
different statistics, such as heavy-tailed or 
structured couplings. 

A natural extension is to asymmetric interactions. 
The pinning mechanism is spectral and never needed 
orthogonality, so a fixed point with activity 
still demands $(1-a^*)\lambda=1$ for the rightmost 
eigenvalue. What changes is what the fluctuations 
do once at the edge. The correlations are no longer 
a simple sum over modes, since non-orthogonal 
eigenvectors weight the sum by their mutual 
overlaps~\cite{chen2024}, and whether this modifies 
the scale-free exponent, or the form of the 
correlations altogether, is an open question. 

Finally, it is remarkably parsimonious that a 
single canonical circuit buys both the normalization 
of responses and the criticality of the collective 
dynamics. One prediction follows: disrupting the 
normalization feedback should destroy the scale-free 
correlations, a possible route to the loss of 
criticality observed in pathologies~\cite{zimmern2020, Meisel2012}.

%\medskip

\emph{Acknowledgements}
This work was supported by the National Eye 
Institute (R01-EY035242) and the National 
Institute of Mental Health (R01-MH137669). 
S.M. acknowledges funding from the Simon’s 
center for computational physical chemistry.
K.M. and S.R. contributed equally to this 
work. Correspondence should be addressed 
to F.M. at fm2452@nyu.edu

\bibliography{references}

@article{bernacchia2011, 
year = {2011}, 
title = {{A reservoir of time constants for memory traces in cortical neurons}}, 
author = {Bernacchia, Alberto and Seo, Hyojung and Lee, Daeyeol and Wang, Xiao-Jing}, 
journal = {Nature Neuroscience}, 
issn = {1097-6256}, 
doi = {10.1038/nn.2752}, 
pmid = {21317906}, 
pmcid = {PMC3079398}, 
pages = {366--372}, 
number = {3}, 
volume = {14}, 
}

@article{zeisler2025, 
year = {2025}, 
title = {{Consistent Hierarchies of Single-Neuron Timescales in Mice, Macaques, and Humans}}, 
author = {Zeisler, Zachary R. and Love, Marques and Rutishauser, Ueli and Stoll, Frederic M. and Rudebeck, Peter H.}, 
journal = {The Journal of Neuroscience}, 
issn = {0270-6474}, 
doi = {10.1523/jneurosci.2155-24.2025}, 
pmid = {40180571}, 
pmcid = {PMC12060611}, 
number = {19}, 
volume = {45}
}

@article{runyan2017, 
year = {2017}, 
title = {{Distinct timescales of population coding across cortex}}, 
author = {Runyan, Caroline A. and Piasini, Eugenio and Panzeri, Stefano and Harvey, Christopher D.}, 
journal = {Nature}, 
issn = {0028-0836}, 
doi = {10.1038/nature23020}, 
pmid = {28723889}, 
pmcid = {PMC5859334}, 
pages = {92--96}, 
number = {7665}, 
volume = {548}, 
}

@article{zeraati2026, 
year = {2026}, 
title = {{Neural timescales from a computational perspective}}, 
author = {Zeraati, Roxana and Levina, Anna and Macke, Jakob H. and Gao, Richard}, 
journal = {Nature Neuroscience}, 
issn = {1097-6256}, 
doi = {10.1038/s41593-026-02343-8}, 
pmid = {42393348}, 
pages = {1534--1547}, 
number = {7}, 
volume = {29}
}

@article{huang2017, 
year = {2017}, 
title = {{Once upon a (slow) time in the land of recurrent neuronal networks…}}, 
author = {Huang, Chengcheng and Doiron, Brent}, 
journal = {Current Opinion in Neurobiology}, 
issn = {0959-4388}, 
doi = {10.1016/j.conb.2017.07.003}, 
pmid = {28756341}, 
pmcid = {PMC12038865}, 
pages = {31--38}, 
volume = {46}
}

@article{chen2024, 
year = {2024}, 
title = {{Searching for long timescales without fine tuning}}, 
author = {Chen, Xiaowen and Bialek, William}, 
journal = {Physical Review E}, 
issn = {2470-0045}, 
doi = {10.1103/physreve.110.034407}, 
pmid = {39425360}, 
pages = {034407}, 
number = {3}, 
volume = {110}, 
}

@article{meshulam2019, 
year = {2019}, 
title = {{Coarse Graining, Fixed Points, and Scaling in a Large Population of Neurons}}, 
author = {Meshulam, Leenoy and Gauthier, Jeffrey L. and Brody, Carlos D. and Tank, David W. and Bialek, William}, 
journal = {Physical Review Letters}, 
issn = {0031-9007}, 
doi = {10.1103/physrevlett.123.178103}, 
pmid = {31702278}, 
pmcid = {PMC7335427}, 
eprint = {1809.08461},  
pages = {178103}, 
number = {17}, 
volume = {123}, 
}

@article{he2014, 
year = {2014}, 
title = {{Scale-free brain activity: past, present, and future}}, 
author = {He, Biyu J.}, 
journal = {Trends in Cognitive Sciences}, 
issn = {1364-6613}, 
doi = {10.1016/j.tics.2014.04.003}, 
pmid = {24788139}, 
pmcid = {PMC4149861}, 
pages = {480--487}, 
number = {9}, 
volume = {18}
}

@article{seung1996, 
year = {1996}, 
title = {{How the brain keeps the eyes still}}, 
author = {Seung, H. S.}, 
journal = {Proceedings of the National Academy of Sciences}, 
issn = {0027-8424}, 
doi = {10.1073/pnas.93.23.13339}, 
pmid = {8917592}, 
pmcid = {PMC24094}, 
pages = {13339--13344}, 
number = {23}, 
volume = {93}, 
}

@article{major2004, 
year = {2004}, 
title = {{Persistent neural activity: prevalence and mechanisms}}, 
author = {Major, Guy and Tank, David}, 
journal = {Current Opinion in Neurobiology}, 
issn = {0959-4388}, 
doi = {10.1016/j.conb.2004.10.017}, 
pmid = {15582368}, 
pages = {675--684}, 
number = {6}, 
volume = {14}
}

@article{beggs2003, 
year = {2003}, 
title = {{Neuronal Avalanches in Neocortical Circuits}}, 
author = {Beggs, John M. and Plenz, Dietmar}, 
journal = {The Journal of Neuroscience}, 
issn = {0270-6474}, 
doi = {10.1523/jneurosci.23-35-11167.2003}, 
pmid = {14657176}, 
pmcid = {PMC6741045}, 
pages = {11167--11177}, 
number = {35}, 
volume = {23}, 
}

@article{plenz2021, 
year = {2021}, 
title = {{Self-Organized Criticality in the Brain}}, 
author = {Plenz, Dietmar and Ribeiro, Tiago L. and Miller, Stephanie R. and Kells, Patrick A. and Vakili, Ali and Capek, Elliott L.}, 
journal = {Frontiers in Physics}, 
doi = {10.3389/fphy.2021.639389}, 
pages = {639389}, 
volume = {9}, 
}

@article{ShewPlenz2013, author={Shew, Woodrow L. and Plenz, Dietmar},
  title={The functional benefits of criticality in the cortex},
  journal={The Neuroscientist}, volume={19}, pages={88--100}, year={2013},
  doi={10.1177/1073858412445487}}

@article{gautam2015, 
year = {2015}, 
title = {{Maximizing Sensory Dynamic Range by Tuning the Cortical State to Criticality}}, 
author = {Gautam, Shree Hari and Hoang, Thanh T. and McClanahan, Kylie and Grady, Stephen K. and Shew, Woodrow L.}, 
journal = {PLoS Computational Biology}, 
issn = {1553-734X}, 
doi = {10.1371/journal.pcbi.1004576}, 
pmid = {26623645}, 
pmcid = {PMC4666488}, 
pages = {e1004576}, 
number = {12}, 
volume = {11}
}

@article{kinouchi2006, 
year = {2006}, 
title = {{Optimal dynamical range of excitable networks at criticality}}, 
author = {Kinouchi, Osame and Copelli, Mauro}, 
journal = {Nature Physics}, 
issn = {1745-2473}, 
doi = {10.1038/nphys289}, 
eprint = {q-bio/0601037}, 
pages = {348--351}, 
number = {5}, 
volume = {2}
}

@article{ganguli2008, 
year = {2008}, 
title = {{Memory traces in dynamical systems}}, 
author = {Ganguli, Surya and Huh, Dongsung and Sompolinsky, Haim}, 
journal = {Proceedings of the National Academy of Sciences}, 
issn = {0027-8424}, 
doi = {10.1073/pnas.0804451105}, 
pmid = {19020074}, 
pmcid = {PMC2596211}, 
pages = {18970--18975}, 
number = {48}, 
volume = {105}, 
}

@article{zimmern2020, 
year = {2020}, 
title = {{Why Brain Criticality Is Clinically Relevant: A Scoping Review}}, 
author = {Zimmern, Vincent}, 
journal = {Frontiers in Neural Circuits}, 
doi = {10.3389/fncir.2020.00054}, 
pmid = {32982698}, 
pmcid = {PMC7479292}, 
pages = {54}, 
volume = {14}
}

@article{Meisel2012,
  author  = {Meisel, Christian and Storch, Alexander and Hallmeyer-Elgner, Susanne and Bullmore, Ed and Gross, Thilo},
  title   = {Failure of Adaptive Self-Organized Criticality during Epileptic Seizure Attacks},
  journal = {PLoS Computational Biology},
  volume  = {8},
  number  = {1},
  pages   = {e1002312},
  year    = {2012},
  doi     = {10.1371/journal.pcbi.1002312}
}

@article{zenari2026, 
year = {2026}, 
title = {{Topological Origin of the Diversity of Timescales in Recurrent Neural Circuits}}, 
author = {Zenari, Marco and Taffarello, Luca and Mazzucato, Luca and Maritan, Amos and Suweis, Samir}, 
journal = {arXiv}, 
doi = {10.48550/arxiv.2603.04149}, 
eprint = {2603.04149}, 
}

@article{rajan2006, 
year = {2006}, 
title = {{Eigenvalue Spectra of Random Matrices for Neural Networks}}, 
author = {Rajan, Kanaka and Abbott, L. F.}, 
journal = {Physical Review Letters}, 
issn = {0031-9007}, 
doi = {10.1103/physrevlett.97.188104}, 
pmid = {17155583}, 
pages = {188104}, 
number = {18}, 
volume = {97}
}

@article{bienenstock1982, 
year = {1982}, 
title = {{Theory for the development of neuron selectivity: orientation specificity and binocular interaction in visual cortex}}, 
author = {Bienenstock, EL and Cooper, LN and Munro, PW}, 
journal = {The Journal of Neuroscience}, 
issn = {0270-6474}, 
doi = {10.1523/jneurosci.02-01-00032.1982}, 
pmid = {7054394}, 
pmcid = {PMC6564292}, 
pages = {32--48}, 
number = {1}, 
volume = {2}, 
}

@article{hengen2025, 
year = {2025}, 
title = {{Is criticality a unified setpoint of brain function?}}, 
author = {Hengen, Keith B. and Shew, Woodrow L.}, 
journal = {Neuron}, 
issn = {0896-6273}, 
doi = {10.1016/j.neuron.2025.05.020}, 
pmid = {40555236}, 
pmcid = {PMC12374783}, 
pages = {2582--2598.e2}, 
number = {16}, 
volume = {113}
}

@article{yellin2025, 
year = {2025}, 
title = {{Adaptive proximity to criticality underlies amplification of ultra-slow fluctuations during free recall}}, 
author = {Yellin, Dovi and Siegel, Noam and Malach, Rafael and Shriki, Oren}, 
journal = {PLOS Computational Biology}, 
issn = {1553-734X}, 
doi = {10.1371/journal.pcbi.1013528}, 
pmid = {41150693}, 
pmcid = {PMC12578353}, 
pages = {e1013528}, 
number = {10}, 
volume = {21}
}

@article{meisel2017, 
year = {2017}, 
title = {{Decline of long-range temporal correlations in the human brain during sustained wakefulness}}, 
author = {Meisel, Christian and Bailey, Kimberlyn and Achermann, Peter and Plenz, Dietmar}, 
journal = {Scientific Reports}, 
doi = {10.1038/s41598-017-12140-w}, 
pmid = {28928479}, 
pmcid = {PMC5605531}, 
pages = {11825}, 
number = {1}, 
volume = {7}
}

@article{Magnasco2009,
  author  = {Magnasco, Marcelo O. and Piro, Oreste and Cecchi, Guillermo A.},
  title   = {Self-Tuned Critical Anti-Hebbian Networks},
  journal = {Physical Review Letters},
  volume  = {102},
  pages   = {258102},
  year    = {2009},
  doi     = {10.1103/PhysRevLett.102.258102}
}

@article{krishnamurthy2022, 
year = {2022}, 
title = {{Theory of Gating in Recurrent Neural Networks}}, 
author = {Krishnamurthy, Kamesh and Can, Tankut and Schwab, David J.}, 
journal = {Physical Review X}, 
issn = {2160-3308}, 
doi = {10.1103/physrevx.12.011011}, 
pmid = {36545030}, 
pmcid = {PMC9762509}, 
pages = {011011}, 
number = {1}, 
volume = {12}
}

@article{can2025, 
year = {2025}, 
title = {{Emergence of Robust Memory Manifolds}}, 
author = {Can, Tankut and Krishnamurthy, Kamesh}, 
journal = {PRX Life}, 
doi = {10.1103/prxlife.3.023006}, 
pages = {023006}, 
number = {2}, 
volume = {3}, 
}

@article{Heeger1992,
  author  = {Heeger, David J.},
  title   = {Normalization of cell responses in cat striate cortex},
  journal = {Visual Neuroscience},
  volume  = {9},
  pages   = {181--197},
  year    = {1992},
  doi     = {10.1017/S0952523800009640}
}

@article{CarandiniHeeger2012,
  author  = {Carandini, Matteo and Heeger, David J.},
  title   = {Normalization as a canonical neural computation},
  journal = {Nature Reviews Neuroscience},
  volume  = {13},
  pages   = {51--62},
  year    = {2012},
  doi     = {10.1038/nrn3136}
}

@article{heeger2019oscillatory,
  title={Oscillatory recurrent gated neural integrator circuits (ORGaNICs), a unifying theoretical framework for neural dynamics},
  author={Heeger, David J and Mackey, Wayne E},
  journal={Proceedings of the National Academy of Sciences},
  volume={116},
  number={45},
  pages={22783--22794},
  year={2019},
  publisher={National Academy of Sciences}
}

@article{atallah2012, 
year = {2012}, 
title = {{Parvalbumin-Expressing Interneurons Linearly Transform Cortical Responses to Visual Stimuli}}, 
author = {Atallah, Bassam V. and Bruns, William and Carandini, Matteo and Scanziani, Massimo}, 
journal = {Neuron}, 
issn = {0896-6273}, 
doi = {10.1016/j.neuron.2011.12.013}, 
pmid = {22243754}, 
pmcid = {PMC3743079}, 
pages = {159--170}, 
number = {1}, 
volume = {73}
}

@article{ma2019, 
year = {2019}, 
title = {{Cortical Circuit Dynamics Are Homeostatically Tuned to Criticality In Vivo}}, 
author = {Ma, Zhengyu and Turrigiano, Gina G. and Wessel, Ralf and Hengen, Keith B.}, 
journal = {Neuron}, 
issn = {0896-6273}, 
doi = {10.1016/j.neuron.2019.08.031}, 
pmid = {31601510}, 
pmcid = {PMC6934140}, 
pages = {655--664.e4}, 
number = {4}, 
volume = {104}
}

@article{morone2026stabilization,
  title={Stabilization of recurrent neural networks through divisive normalization},
  author={Morone, Flaviano and Rawat, Shivang and Heeger, David J and Martiniani, Stefano},
  journal={Proceedings of the National Academy of Sciences},
  volume={123},
  number={30},
  pages={e2601841123},
  year={2026},
  publisher={National Academy of Sciences}
}

@article{perich2025, 
year = {2025}, 
title = {{A neural manifold view of the brain}}, 
author = {Perich, Matthew G. and Narain, Devika and Gallego, Juan A.}, 
journal = {Nature Neuroscience}, 
issn = {1097-6256}, 
doi = {10.1038/s41593-025-02031-z}, 
pmid = {40721675}, 
pages = {1582--1597}, 
number = {8}, 
volume = {28}
}

@article{fosque2025, 
year = {2025}, 
title = {{Two views of the brain are reconciled by a unifying principle of maximal information processing}}, 
author = {Fosque, Leandro J. and Shew, Woodrow L. and Ching, ShiNung and Hengen, Keith B.}, 
journal = {bioRxiv}, 
doi = {10.1101/2025.11.25.690580}, 
pmid = {41394724}, 
pmcid = {PMC12697366}, 
pages = {2025.11.25.690580}
}

\end{document}